\documentclass[12pt]{article}
\usepackage{putex}
\usepackage{graphicx}
\usepackage{amssymb,amsmath}

\begin{document}
\preprint{PUPT-2427}

\institution{PU}{Joseph Henry Laboratories, Princeton University, Princeton, NJ 08544, USA}

\title{Complex deformations of Bjorken flow}

\authors{Steven S.~Gubser}

\abstract{Through a complex shift of the time coordinate, a modification of Bjorken flow is introduced which interpolates between a glasma-like stress tensor at forward rapidities and Bjorken-like hydrodynamics around mid-rapidity.  A Landau-like full-stopping regime is found at early times and rapidities not too large.  Approximate agreement with BRAHMS data on the rapidity distribution of produced particles at top RHIC energies can be achieved if the complex shift of the time coordinate is comparable to the inverse of the saturation scale.  The form of the stress tensor follows essentially from symmetry considerations, and it can be expressed in closed form.}

\date{October 2012}

\maketitle

\section{Outline of the construction}

The central idea of Bjorken flow \cite{Bjorken:1982qr} is that there is approximate boost invariance in the direction of the beamline for the dynamics near mid-rapidity.  Particles that form near mid-rapidity are assumed to do so at some definite, boost-invariant proper time $\tau_{\rm form} \approx 1\,{\rm fm}/c$, where $\tau = \sqrt{t^2-x_3^2}$.  Assuming that all motion is in the beamline direction, the individual four-velocities of the produced particles can be deduced to have the form
 \eqn{BjorkenU}{
  u_\mu = \left( -{t \over \tau}, 0, 0, {x_3 \over \tau} \right) \,.
 }
To the extent that one may use a hydrodynamical description, the boost invariance also constrains the local energy density $\epsilon$: it can depend on $t$ and $x_3$ only through the combination $\tau$.  If one assumes translational and rotational invariance in the collision plane, then $\epsilon$ cannot depend on $x_1$ or $x_2$.  Thus $\epsilon = \epsilon(\tau)$.  One can obtain explicit expressions for $\epsilon(\tau)$ by assuming a specific form for the stress tensor.  For example, let's choose
 \eqn{TmnInviscid}{
  T_{\mu\nu} = \epsilon u_\mu u_\nu + {\epsilon \over 3} (g_{\mu\nu} + u_\mu u_\nu) \,,
 }
corresponding to inviscid, conformal hydrodynamics.  Then one immediately finds
 \eqn{FoundEpsilon}{
  \epsilon(\tau) = {\epsilon_0 \over \tau^{4/3}} \,,
 }
where $\epsilon_0$ is a constant.

A trivial modification of Bjorken flow is to set
 \eqn{uCform}{
  u^{\bf C}_\mu &= \left( -{t+\mathfrak{t}_3 \over \sqrt{(t+\mathfrak{t}_3)^2 - x_3^2}}, 0, 0, 
    {x_3 \over \sqrt{(t+\mathfrak{t}_3)^2 - x_3^2}} \right)  \cr
  \epsilon^{\bf C} &= {\epsilon^{\bf C}_0 \over ((t+\mathfrak{t}_3)^2-x_3^2)^{2/3}}  \cr
  T^{\bf C}_{\mu\nu} &= \epsilon^{\bf C} u^{\bf C}_\mu u^{\bf C}_\nu + 
    {\epsilon^{\bf C} \over 3} (g_{\mu\nu} + u^{\bf C}_\mu u^{\bf C}_\nu) \,,
 }
where $\mathfrak{t}_3$ is a constant.  If $\mathfrak{t}_3$ is real, then we have simply translated Bjorken flow in time.  If $\mathfrak{t}_3$ is complex, we have something new, and all the quantities with a superscript ${\bf C}$ become complex.  The complexified stress tensor $T^{\bf C}_{\mu\nu}$ still obeys the conservation equations $\nabla^\mu T^{\bf C}_{\mu\nu} = 0$, and because these are linear equations, a conserved stress tensor with all components real can be obtained by defining
 \eqn{TmnRe}{
  T_{\mu\nu} \equiv \Re T^{\bf C}_{\mu\nu} \,.
 }
$T_{\mu\nu}$ as identified in \eno{TmnRe} will not in general satisfy the inviscid hydrodynamic constitutive relations.

\begin{figure}[t]
  \centerline{\includegraphics[width=4in]{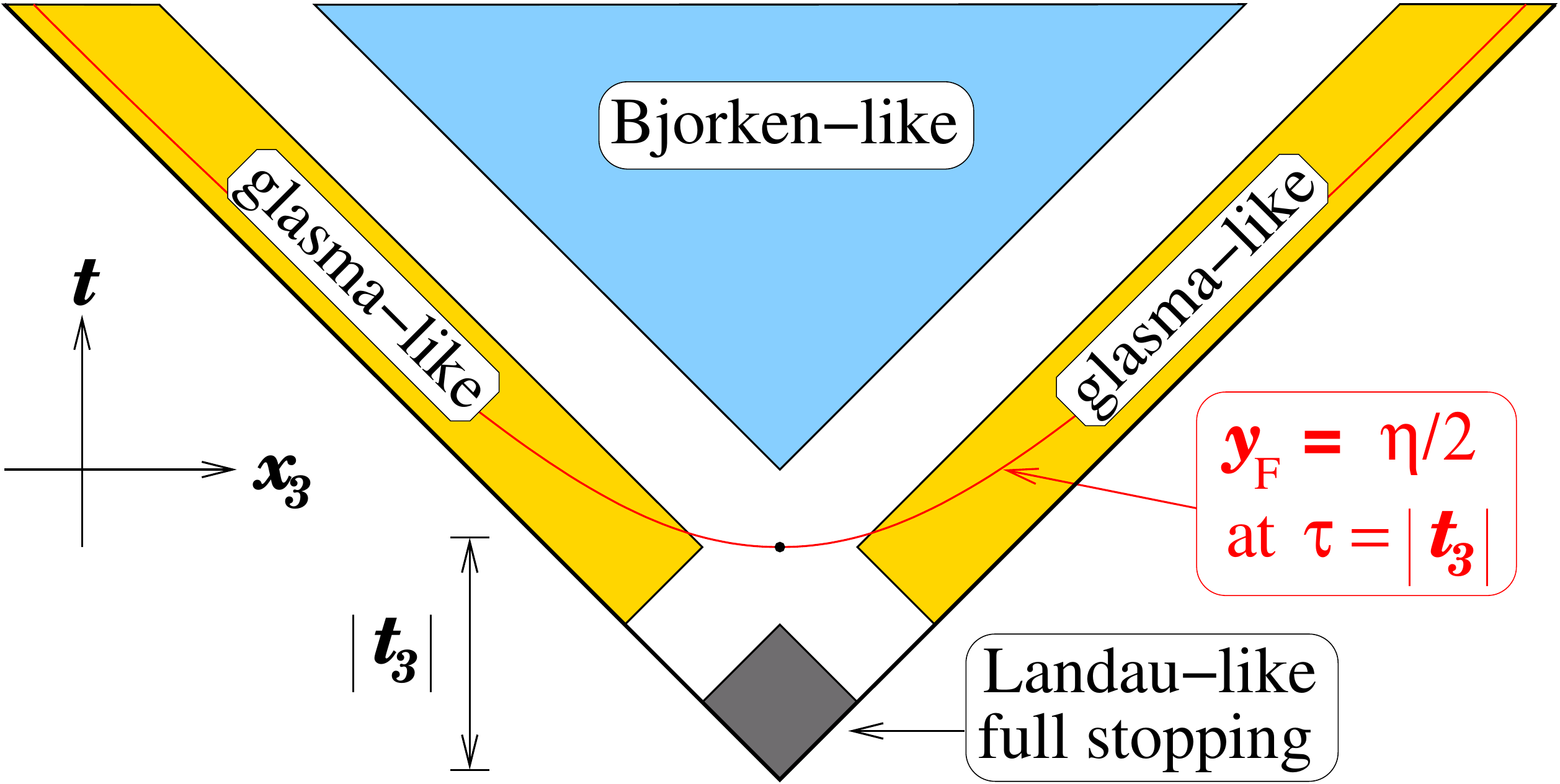}}
  \caption{A simplified cartoon of the rapidity structure of the complex deformation of Bjorken flow indicated in \eno{uCform}-\eno{TmnRe}, with $\arg\epsilon^{\bf C} = \pi/3$ and $\arg\mathfrak{t}_3 = \pi/2$.}\label{CartoonWedge}
 \end{figure}

It may seem that \eno{uCform}-\eno{TmnRe} are an unmotivated and unpromising line of attack on the problem of describing the rapidity structure of heavy ion collisions.  However,  provided the phase of $\epsilon^{\bf C}$ is chosen correctly (namely $\arg \epsilon^{\bf C} = \pi/3$ when $\arg\mathfrak{t}_3 = \pi/2$), an appealing spacetime picture emerges in the forward lightcone, as illustrated in figure~\ref{CartoonWedge}.  At early times and rapidities not too large, there is a full-stopping region reminiscent of the Landau model.  Although the hydrodynamic constitutive relations hold to good accuracy in this region, they do not hold uniformly in its causal future.  Instead, one recovers the hydrodynamic constitutive relations, and Bjorken flow, asymptotically at late proper times $\tau$ with rapidity held fixed; but at forward rapidities one obtains a glasma-like form of the stress tensor, with longitudinal pressure nearly equal to minus the energy density.  Interestingly, at $\tau = |\mathfrak{t}_3|$, there is a very simple relation between the rapidity $y_F$ of the fluid and the pseudorapidity:
 \eqn{yFRelation}{
  y_F = {\eta \over 2} \qquad\hbox{when}\quad \tau = |\mathfrak{t}_3| \,.
 }
This is to be compared with the relation $y_F = \eta$ for Bjorken flow.

The organization of the rest of this paper is as follows.  In section~\ref{POSITIVITY} I explain how the phase of $\epsilon^{\bf C}$ is fixed by positive energy considerations.  In section~\ref{FEATURES} I explain the main features of the rapidity structure of the flow, providing in particular a more quantitative version of the diagram in figure~\ref{CartoonWedge}.  In section~\ref{HADRONIZE} I consider a simplified version of hadronization and exhibit a comparison of the predicted rapidity profile with data from the BRAHMS experiment.  I conclude in section~\ref{DISCUSSION} with a discussion of symmetry groups and local entropy production.

\section{Positive energy constraints}
\label{POSITIVITY}

In addition to the conservation equations, a sensible stress tensor must satisfy some sort of positive energy constraints.  The conditions we will apply are that the Landau frame can be defined throughout the future light-wedge of the collision plane, and that the energy density in the Landau frame is positive throughout this wedge.

To pass from the laboratory frame to the Landau frame, we must apply a boost which makes the stress tensor diagonal.\footnote{For a more careful discussion of how to define Landau frame, see for example \cite{Bhattacharya:2011tra}.}  This boost is in the $x_3$ direction, because the only non-diagonal components of $T_{\mu\nu}$ are $T_{03} = T_{30}$.  Suppressing the $x_1$ and $x_2$ directions for brevity, we may express
 \eqn[c]{TLForm}{
  T_{\mu\nu} = \begin{pmatrix} T_{00} & T_{03} \\ T_{03} & T_{33} \end{pmatrix} \qquad
  \Lambda^\mu{}_\nu = 
    \begin{pmatrix} \cosh y_F & \sinh y_F \\ \sinh y_F & \cosh y_F \end{pmatrix}  \cr
  T^L_{\mu\nu} = \Lambda^\alpha{}_\mu \Lambda^\beta{}_\nu T_{\alpha\beta} =
    \begin{pmatrix} \epsilon^L & 0 \\ 0 & p_3^L \end{pmatrix} \,,
 }
where $\epsilon^L$ is the Landau frame energy density and $p_3^L$ is the Landau frame longitudinal pressure.  The transverse pressure $p_T = p_1 = p_2$ is the same in the Landau frame as it is in our original frame.  It is easy to show from \eno{TLForm} that
 \eqn{yFForm}{
  y_F &= -{1 \over 2} \tanh^{-1} {2 T_{03} \over T_{00} + T_{33}}  \cr
  \epsilon^L &= {T_{00} - T_{33} + 
     \sqrt{(T_{00}-2T_{03}+T_{33})(T_{00}+2T_{03}+T_{33})} \over 2}  \cr
  p_3^L &= {-T_{00} + T_{33} + 
     \sqrt{(T_{00}-2T_{03}+T_{33})(T_{00}+2T_{03}+T_{33})} \over 2} \,.
 }
The Landau frame exists precisely when $y_F$ as defined in \eno{yFForm} is real.  I claim that in order for the Landau frame to exist throughout the future light-cone of the collision plane, one must set
 \eqn{epsilonPhase}{
  \epsilon_0^{\bf C} = i^{2/3} \epsilon_0^{\bf R} \,,
 }
where $\epsilon_0^{\bf R}$ is real.  Furthermore, $\epsilon^L > 0$ everywhere in the future light-cone provided $\epsilon_0^{\bf R} > 0$.  A full demonstration of these claims is tedious, but I will give sufficient indications here to show that no other phase than the one in \eno{epsilonPhase} will suffice.  To this end, let's set
 \eqn{SpecialValues}{
  \epsilon_0^{\bf C} = i^{2/3} e^{i\theta} \qquad \mathfrak{t}_3 = i \qquad
   \tau = \sqrt{t^2 - x_3^2} = 1 \,,
 }
where without loss of generality we can limit $\arg \theta \in (-\pi/2,\pi/2)$ by allowing $\epsilon_0^{\bf R}$ to be a signed real quantity (anticipating that the constraint $\epsilon_0^{\bf R} > 0$ will emerge from later arguments).  Straightforward computation now leads to
 \eqn{yFSpecial}{
  \tanh 2y_F = {-2 T_{03} \over T_{00} + T_{33}} = 
    {\sqrt{t^2-1} \over t} \left( 1 + {\sin\theta \over t \cos\theta + (t^2-1) \sin\theta} 
     \right) \,,
 }
It should be borne in mind that \eno{yFSpecial} holds only with the special values \eno{SpecialValues}, in particular $\tau = 1$.  If $\sin\theta \neq 0$, then by expanding \eno{yFSpecial} at large $t$ one can see immediately that the right hand side is greater than $1$ when $t$ is sufficiently large.  Thus $\theta=0$, and \eno{yFSpecial} simplifies dramatically to $y_F = \eta/2$, where
 \eqn{PseudoRapidity}{
  \eta = \tanh^{-1} {x_3 \over t}
 }
is the spacetime pseudorapidity.  This result, previously quoted as \eno{yFRelation}, contrasts with the result $y_F = \eta$ for standard Bjorken flow and provides some preliminary indication that the flow is more focused near mid-rapidity.  It is also useful to note that $y_F = 0$ when $\eta = 0$ for all $t$, and upon setting $\theta = 0$ one finds
 \eqn{epsilonMid}{
  T_{00} = \epsilon^L = 3 p_3^L = 3 p_T = \Re {i^{2/3} \epsilon_0^{\bf R} \over 
    (t + \mathfrak{t}_3)^{4/3}} \qquad\hbox{when}\quad x_3 = 0 \,.
 }
So the ideal hydrodynamic constitutive relations are satisfied precisely at mid-rapidity, and $T_{00} \approx \epsilon_0^{\bf R} / 2 t^{4/3}$ at mid-rapidity for $t \gg |\mathfrak{t}_3|$, which is the same functional dependence as for Bjorken flow.  Thus $\epsilon_0^{\bf R} > 0$, as promised.

\section{Features of the rapidity dependence}
\label{FEATURES}

 \begin{figure}
  \centerline{\includegraphics[width=4in]{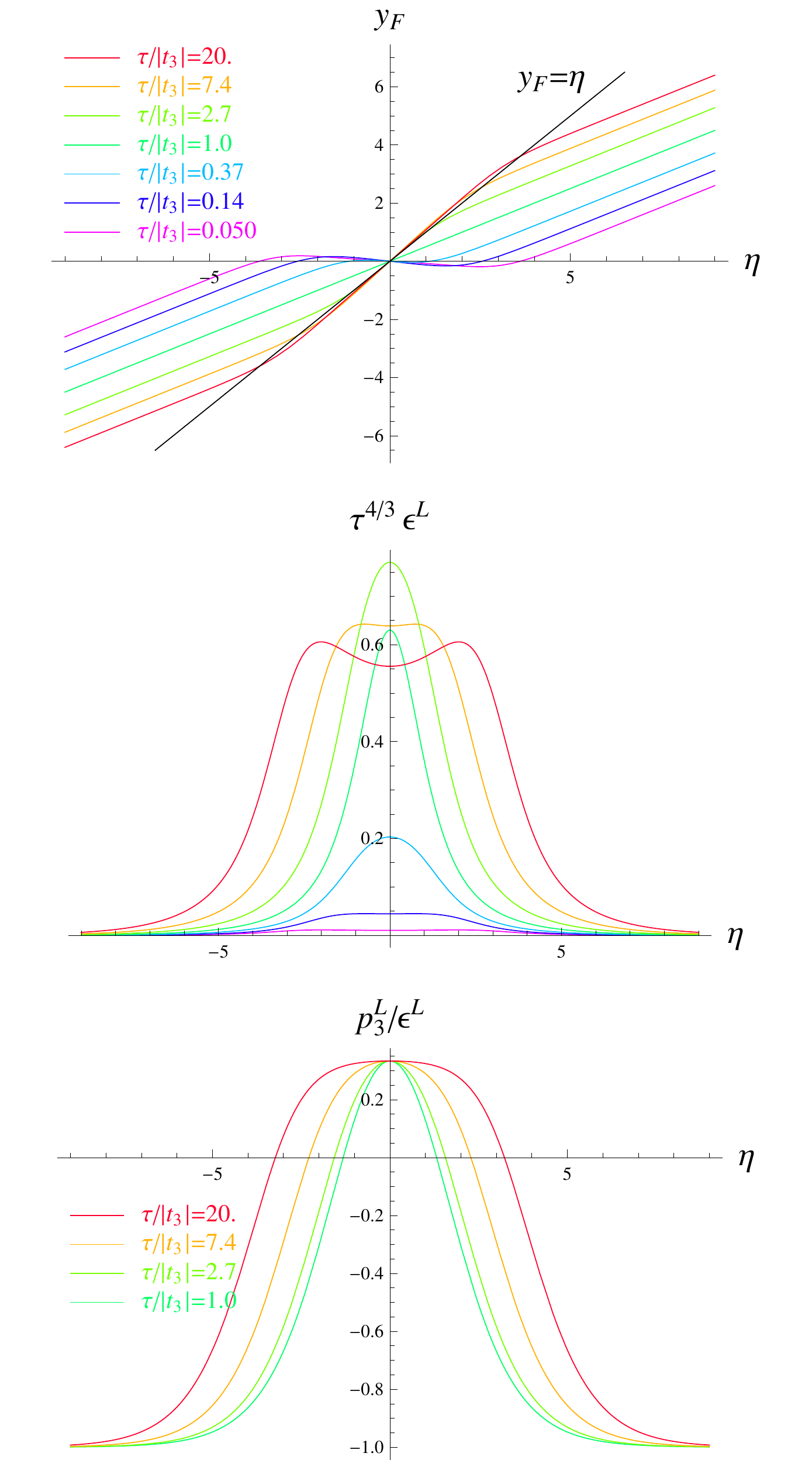}}
  \caption{Top: The fluid rapidity versus spacetime pseudorapidity, at several different values of Bjorken time, for $\mathfrak{t}_3 = i$.  Middle: The Landau frame energy density, scaled by $\tau^{-4/3}$, versus spacetime pseudorapidity, at several different values of Bjorken time.  Bottom: The ratio of longitudinal pressure to energy density, in the Landau frame, versus spacetime pseudorapidity, at several different values of Bjorken time.  The ratio $p_3^L / \epsilon^L$ is not shown for $\tau < 1$ because it happens to be identical for $\tau$ and $|\mathfrak{t}_3|^2/\tau$.}\label{FlowFeatures}
 \end{figure}
\begin{figure}
  \centerline{\includegraphics[width=4in]{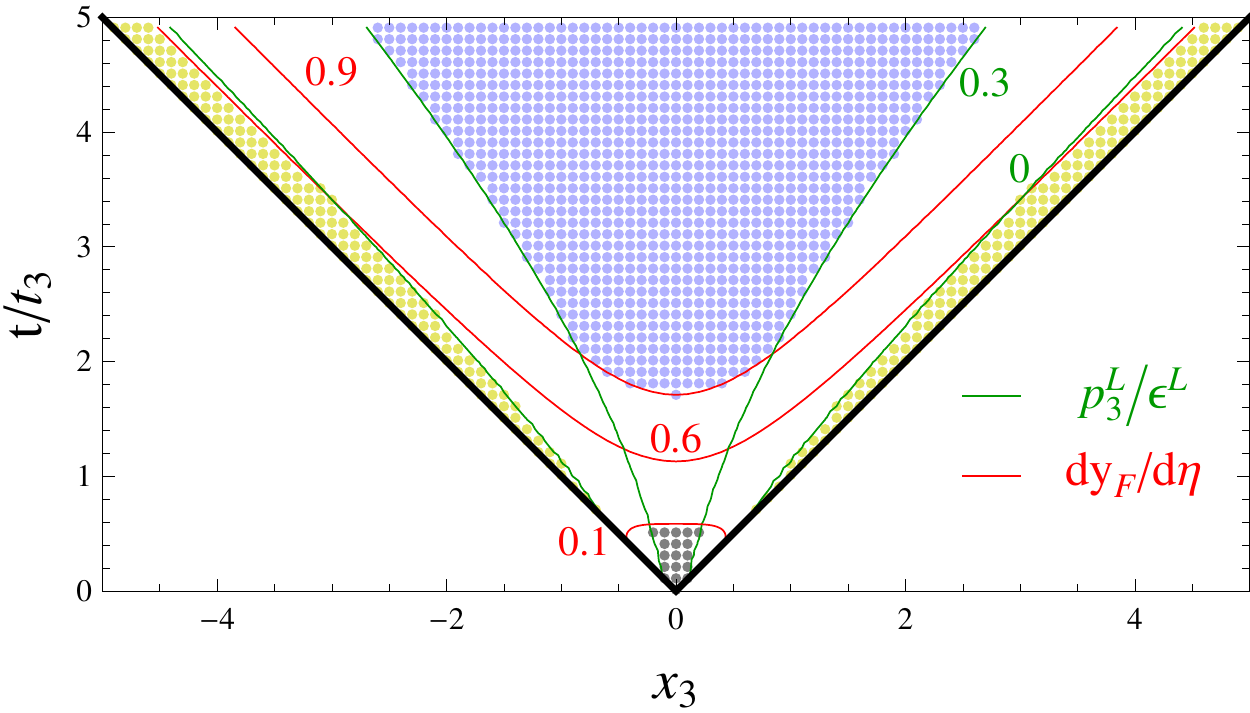}}\vskip0.5in
  \caption{Contours of constant $p_3^L/\epsilon^L$ (green) and $dy_F/d\eta$ (red) demarcate regions associated with Bjorken-like flow (blue), Landau-like full-stopping (grey) and glasma-like behavior (gold).}\label{ShadedWedge}
 \end{figure}

Up to an overall rescaling of time, $\mathfrak{t}_3 = i$ is the unique choice for the type of flow we are interested in.  As argued above, up to an overall rescaling of energy, $\epsilon_0^{\bf C} = i^{2/3}$ is the unique choice for a physically sensible stress tensor.  Thus, through \eno{uCform}, \eno{TmnRe}, and \eno{yFForm}, I have constructed an essentially unique stress tensor.  Its main features, exhibited in figures~\ref{FlowFeatures}-\ref{ShadedWedge}, may be summarized as follows:
 \begin{itemize}
  \item There is a region at late times in which $y_F \approx \eta$ and $p_3^L / \epsilon \approx 1/3$: Bjorken flow is approximately recovered.  To define more precisely where this happens, I note that
 \eqn{BjorkenRegion}{
  {p_3^L \over \epsilon^L} > 0.3 \qquad&\hbox{when}\qquad
   |\eta| \lsim -1.4 + \log {\tau \over |\mathfrak{t}_3|}  \cr
  {dy_F \over d\eta} > 0.9 \qquad&\hbox{when}\qquad 
   |\eta| \lsim \log {\tau \over |\mathfrak{t}_3|} \,.
 }
The energy density and pressure become broader in pseudorapidity, and the energy density acquires a characteristic double-hump structure for $t \gsim 10 |\mathfrak{t}_3|$.  It is interesting that $y_F$ attains values slightly larger than $\eta$ in the Bjorken region (by roughly $10\%$).  $y_F$ then approaches $\eta$ from {\it above} as $\tau \to \infty$ at fixed $\eta$.  Because of this overshoot, it is perhaps better to describe the region where the conditions \eno{BjorkenRegion} hold as ``Bjorken-like.''
  \item There is a region at early times in which the fluid rapidity $y_F$ is small and the ratio $p_3^L / \epsilon^L$ is close to $1/3$.  This is reminiscent of full stopping in the Landau model, so I will refer to the region in question as the full-stopping region.  Because $y_F = 0$ exactly at mid-rapidity for symmetry reasons, it is better to use the smallness of $dy_F/d\eta$ (at fixed $\tau$) to define the full-stopping region.  I find that
 \eqn{FullStoppingRegion}{
  {dy_F \over d\eta} < 0.1 \quad\hbox{and}\quad {p_3^L \over \epsilon^L} > 0.3
   \qquad\hbox{when}\quad {t \over \mathfrak{t}_3} \lsim 0.6 \quad\hbox{and}\quad
    {x_3 \over |\mathfrak{t}_3|} \lsim 0.2 \,.
 }
It is probably appropriate to think of the fluid as far from equilibrium in the full-stopping region, because although the fluid is close to satisfying the inviscid hydrodynamic constitutive relations there, it ceases to do so in much of the causal future of the full-stopping region: in particular, in the glasma-like region to be described next.
  \item At extreme forward rapidities (and for proper times not too small) the fluid rapidity is close to satisfying the curious relation $dy_F/d\eta = 1/2$---a relation which, as explained around \eno{yFSpecial}, is exactly satisfied for all rapidities when $\tau = |\mathfrak{t}_3|$.  In a similar region, I find $p_3^L < 0$, with $p_3^L \approx -\epsilon^L$ as the pseudorapidity becomes large.  This is the same as the stress tensor of purely longitudinal electric and/or magnetic fields, as considered in glasma accounts of pre-thermalization dynamics: see for example \cite{Lappi:2006fp}.  At late times,
 \eqn{GlasmaRegion}{
  {p_3^L \over \epsilon^L} < 0 \qquad&\hbox{when}\qquad
   |\eta| \gsim 0.2 + \log {\tau \over |\mathfrak{t}_3|}  \cr
  {dy_F \over d\eta} < 0.6 \qquad&\hbox{when}\qquad 
   |\eta| \gsim 0.9 + \log {\tau \over |\mathfrak{t}_3|} \,.
 }
 \end{itemize}
A visual summary of the bulleted points above can be found in figure~\ref{ShadedWedge}, which may be seen as a more quantitative version of figure~\ref{CartoonWedge}.

\section{A simple version of hadronization}
\label{HADRONIZE}

The presence of a Bjorken plateau in the energy density which widens as one proceeds toward late times is, at least qualitatively, a phenomenologically attractive feature.  To be more quantitative, one should ask what sort of spectrum of hadrons one gets from isothermal freezeout.  Using a conformal equation of state all the way to freezeout seems excessively naive, so let's use the equation of state sometimes referred to as EOS Q \cite{Kolb:2003dz}.  In this model,
 \eqn{EOSQ}{
  p = {\epsilon - 4B \over 3} \qquad\hbox{for}\qquad T > T_c \,,
 }
where $B$ is the bag constant and $T_c$ is the deconfinement temperature.  Using the first law of thermodynamics one can deduce from \eno{EOSQ} that
 \eqn{EOSQT}{
  \epsilon = {\pi^2 \over 30} g_* T^4 + B \quad\hbox{and}\quad 
   p = {\pi^2 \over 90} g_* T^4 - B \qquad\hbox{for}\qquad
   T > T_c \,.
 }
Here $g_*$ is the number of degrees of freedom in the deconfined phase.  EOS Q calls for a strongly first order transition at $T = T_c$, and for $T < T_c$ one assumes $p = 0.15 \epsilon$, which is supposed to describe a Hagedorn spectrum of non-interacting hadronic resonances.  A fairly realistic choice of parameters is
 \eqn{EOSQparams}{
  B = 0.35 \, {{\rm GeV} \over {\rm fm}^3} \qquad
   g_* = 40 \qquad T_c = 164\,{\rm MeV} \,.
 }
It is hard to see how to incorporate an equation of state with a phase transition into the stress tensor construction \eno{TmnRe}.  So let's hadronize at the temperature $T_c$, just before the phase transition occurs.  Usually Cooper-Frye is applied using a lower freezeout temperature, for example $T_{\rm kin} \approx 110\,{\rm MeV}$.\footnote{Hadronization schemes with separate temperatures for chemical and kinetic freezeout typically set the chemical freezeout temperature close to $160\,{\rm MeV}$, close to our $T_c$.  It would be interesting to see whether these more sophisticated hadronization schemes lead to significantly different results for the rapidity distribution.}  The justification for the Cooper-Frye algorithm relies on local equilibration prior to freezeout, so it is dubious to apply it outside the Bjorken plateau.  I will use a freezeout surface running from $\eta = -\log {\tau \over |\mathfrak{t}_3|}$ to $\eta = +\log {\tau \over |\mathfrak{t}_3|}$.  Depending on one's precise definitions, this is roughly the extent of the Bjorken plateau.

By restricting attention to the high-temperature regime \eno{EOSQT}, we can view the stress tensor of EOS Q as that of a CFT plus a positive cosmological constant.  It is perhaps unsurprising that fluid rapidity and temperature (suitably defined) are the same as for a CFT.  The remainder of this paragraph is devoted to a more careful demonstration of this claim.  First note that ordinary Bjorken flow with EOS Q (and no viscosity) has
 \eqn{BjorkenQ}{
  \epsilon_{\rm Q} &= B + {\epsilon_0 \over \tau^{4/3}} = 
    B + \epsilon_{\rm CFT}  \cr
  p_{\rm Q} &= -B + {\epsilon_0 / 3 \over \tau^{4/3}} = 
    -B + p_{\rm CFT} \,.
 }
Because $u_\mu$ in Bjorken flow is completely independent of the equation of state, \eno{BjorkenQ} implies
 \eqn{TmunuQ}{
  (T_{\mu\nu})_{\rm Q} = (T_{\mu\nu})_{\rm CFT} - B g_{\mu\nu} \,.
 }
Passing to a complexified stress tensor, as in \eno{uCform}, with 
 \eqn{epsilonCQ}{
  \epsilon^{\bf C} = B + {\epsilon_0^{\bf C} \over ((t + \mathfrak{t}_3)^2 - x_3^2)^{2/3}}
   \,,
 } 
one sees that $(T_{\mu\nu}^{\bf C})_Q = (T_{\mu\nu}^{\bf C})_{\rm CFT} - B g_{\mu\nu}$, and it follows immediately that the final stress tensor $\Re T_{\mu\nu}^{\bf C}$ obeys the same relation: that is, \eno{TmunuQ} applies unaltered.  Because the term $-B g_{\mu\nu}$ is frame-independent, the boost required to pass to Landau frame is the same whether this term is present or absent.  So $y_F$ is indeed the same for EOS Q as for a CFT.  Temperature is a little trickier because the final stress tensor does not satisfy the hydrodynamic constitutive relations.  The obvious approach is to define
 \eqn{TdefQ}{
  T = \left( {30 \over \pi^2} {\epsilon_L-B \over g_*} \right)^{1/4} \,.
 }
This definition follows from plugging the Landau frame energy density into the first relation in \eno{EOSQT} and solving for the temperature.  $T$ as defined in \eno{TdefQ} evolves identically to the temperature of a CFT, both in Bjorken flow and in  complex deformations of it.

I am going to use a simplified version of the Cooper-Frye expressions for the distribution of outgoing particles $d^3N/dp^3$ with energy $E$: namely
 \eqn{CF}{
  E {d^3 N \over dp^3} = {g \over (2\pi)^3} \int_\Sigma {1 \over 3!} 
    \epsilon_{\mu\nu\rho\sigma} 
    dx^\mu dx^\nu dx^\rho \, p^\sigma e^{u_\lambda p^\lambda / T} \,,
 }
where $u_\mu = (-\cosh y_F,0,0,\sinh y_F)$ in the laboratory reference frame and temperature $T$ is given by \eno{TdefQ}.  (My convention is to sum over indices without regard to ordering, which is the reason we require the explicit $1/3!$ in the integrand.)  The integration surface $\Sigma$ is determined by the equation $T = T_{\rm freezeout}$, where $T_{\rm freezeout}=T_c$ is the freezeout temperature.  The overall factor $g$ refers to the degeneracy of produced particles.  The rapidity of a produced particle can be defined (in the laboratory frame) in terms of its momentum $p^\mu$ as
 \eqn{yDef}{
  y = \tanh^{-1} {p_3 \over E} \,.
 }
Standard manipulations lead to an expression for the rapidity distribution of produced particles:
 \eqn{dNdy}{
  {dN \over dy} = \int dp_1 dp_2 \, E {d^3 N \over dp^3} 
    = \int_\Sigma (q_\tau d\tau + q_\eta d\eta)
 }
where
 \eqn{qForm}{
  q_\tau = Q \sinh(y-\eta) \qquad q_\eta = -Q \tau \cosh(y-\eta)
 }
and
 \eqn{QForm}{
  Q = {g \Vol_\perp \over (2\pi)^2} \int_m^\infty dm_\perp \, m_\perp^2 
     e^{-{m_\perp \over \tilde{T}}} 
    = {g \Vol_\perp \over 2\pi^2} \left( \tilde{T}^3 + m \tilde{T}^2 + 
        {1 \over 2} m^2 \tilde{T} \right) e^{-{m \over \tilde{T}}} \,.
 }
Here $\Vol_\perp$ is the volume in the $x_1$-$x_2$ directions, $m_\perp = \sqrt{p_1^2+p_2^2+m^2}$, and
 \eqn{tildeTDef}{
  \tilde{T} = T \sech(y-y_F) \,.
 }

\begin{figure}
  \centerline{\includegraphics[width=4in]{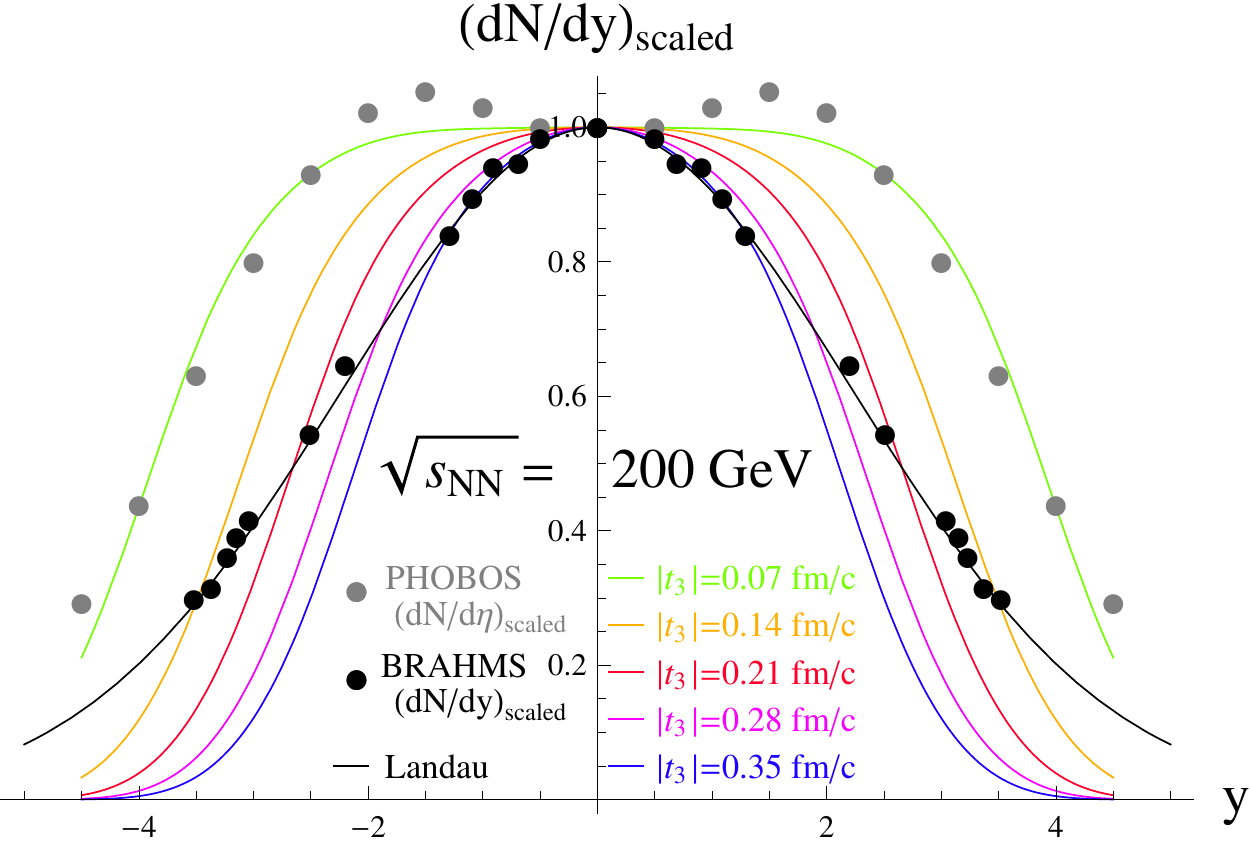}}
  \caption{The rapidity distribution of produced particles in rapidity, normalized to unity at mid-rapidity.  The grey dots are PHOBOS data for $dN/d\eta$, from the lower-right panel of Figure~18 of \cite{Alver:2010ck}.  The black dots are BRAHMS data for $dN/dy$ of positively charged pions, from Figure~4 of \cite{Murray:2004gh}.  The black curve is from the purely hydrodynamic Landau model, as explained in \cite{Murray:2004gh}.}\label{HadronizedFlow}
 \end{figure}

In figure~\ref{HadronizedFlow} I show predictions of $dN/dy$ as compared to data for central RHIC collisions at $\sqrt{s_{NN}} = 200\,{\rm GeV}$.  The main points to note about my parameter choices are:
 \begin{enumerate}
  \item I consider only pions as the outgoing particles, so I use $m=m_\pi \approx 140\,{\rm MeV}$ in \eno{QForm}.
  \item For $|\mathfrak{t}_3|$, I consider multiples of the length $0.07\,{\rm fm}$, which is about half the thickness of the incoming nucleus.
  \item The energy density at $\tau = 0.7\,{\rm fm}$ and $\eta=0$ is set equal to $7.7\,{\rm GeV}/{\rm fm}^3$ in order to approximate the conditions of a central RHIC collision at $\sqrt{s_{NN}} = 200\,{\rm GeV}$ (see for example \cite{Adcox:2004mh}).
  \item I restricted the rapidity range of the freezeout surface to $|\eta| < \log {\tau \over |\mathfrak{t}_3|}$, corresponding approximately to the extent of the Bjorken plateau.
  \item The freezeout surface turns out to be fairly close to isochronous (in Bjorken time $\tau$), with $\tau_{\rm freezeout} \approx 2.3\,{\rm fm}$.  Early freezeout is expected because of the higher-than-typical value of freezeout temperature.
  \item I normalize $dN/dy$ to $1$ at mid-rapidity.  Estimates of the effective $\Vol_\perp$ and inclusion of more hadron species would be needed in order to obtain realistic $dN/dy$ at mid-rapidity.
 \end{enumerate}
Comparison with the BRAHMS data on $dN/dy$ favors $|\mathfrak{t}_3| \approx 0.2\,{\rm fm}$, corresponding to an energy scale of $1\,{\rm GeV}$.  It is gratifying that this is close to the saturation scale $Q_s \sim 1.5\,{\rm GeV}$ for central gold-gold collisions at $\sqrt{s_{NN}} = 200\,{\rm GeV}$.

\section{Discussion}
\label{DISCUSSION}

The key principle underlying Bjorken flow is boost-invariance in the longitudinal direction.  Complexification weakens this principle but does not destroy it entirely.  To see how symmetries behave in the complexification, let's introduce notation for the generators of the Poincar\'e group $ISO(3,1)$:
 \begin{itemize}
  \item Translations, $T_{(\mu)} = \partial_\mu$.  Thus, for example, $T_{(1)}$ is translation in the $x_1$ direction.
  \item Spatial rotations, $R_{(ij)} = x_i \partial_j - x_j \partial_i$.
  \item Boosts, $B_{(i)} = t \partial_i + x_i \partial_t$.
 \end{itemize}
Usually one demands invariance under $B_{(3)}$.  There is essentially one combination of $t$ and $x_3$ that is $B_{(3)}$-invariant, namely $\tau^2 = t^2 - x_3^2$.  If we instead demand invariance under
 \eqn{bDef}{
  b = B_{(3)} + \mathfrak{t}_3 T_{(3)} \,,
 }
then the invariant combination of $t$ and $x_3$ is $(t+\mathfrak{t}_3)^2 - x_3^2$.  It is straightforward to show that ${\cal L}_b u_\mu^{\bf C} = 0$, where ${\cal L}_b$ is the Lie derivative and $u_\mu^{\bf C}$ is defined as in \eno{uCform}.  Likewise, ${\cal L}_b T_{\mu\nu}^{\bf C} = 0$.  When $\mathfrak{t}_3$ is imaginary---the case of interest---$b$ is not a member of the algebra of $ISO(3,1)$, but it is in the complexification of this algebra.  So the complexified stress tensor is invariant under the complexified symmetry.  However, the final form $\Re T_{\mu\nu}^{\bf C}$ is not invariant under $b$ or any obvious modification of it.

Although I have focused exclusively on the case $c_s^2 = 1/3$, it is worth noting that other constant speeds of sound can be treated in essentially the same way.  The relation $y_F = \eta/2$ is recovered at $t=|\mathfrak{t}_3|$ for arbitrary $c_s^2$.  Another way to say this is that the construction goes through for conformal field theories in dimensions other than $3+1$.

Following the definition \eno{TdefQ} of temperature in terms of energy density, and restricting attention to a conformal equation of state, one may define an entropy current
 \eqn{EntropyCurrent}{
  s_\mu = T^3 u_\mu \,.
 }
(There should be an overall constant factor on the right hand side of \eno{EntropyCurrent}, proportional to the number of degrees of freedom, but this factor doesn't affect the discussion to follow.)  When the inviscid hydrodynamic constitutive relations hold, the equations of motion $\nabla_\mu T^{\mu\nu} = 0$ imply conservation of entropy, $\nabla_\mu s^\mu = 0$.  The second law of thermodynamics, applied locally, requires locally increasing entropy, $\nabla_\mu s^\mu \geq 0$.  By direct calculation starting from \eno{uCform}-\eno{TmnRe}, I found that entropy locally increases in the Bjorken-like region, but not in the full-stopping region, and in only a part of the glasma-like region which is not too close to the lightcone.\footnote{More precisely, $\nabla_\mu s^\mu = 0$ for all $\tau$ precisely at mid-rapidity; away from $\eta=0$, $\nabla_\mu s^\mu$ switches signs from negative to positive as $\tau$ increases past an $\eta$-dependent threshold.  This threshold is close to $\tau = |\mathfrak{t}_3|/2$ for near mid-rapidity, and its form at larger rapidities is roughly $\tau = |\mathfrak{t}_3| e^{\eta/3}$.}  Failure of the second law outside the Bjorken-like region seems at first alarming; but what one should conclude is that \eno{EntropyCurrent} is a poor approximation to the entropy current except in the Bjorken-like region.  This makes sense because only in that region is there good reason to think that local equilibration has occurred.

\section*{Acknowledgments}

I thank A.~Yarom for collaboration on the early stages of this work, and for comments on the draft.  I am also indebted to P.~Chesler and E.~Iancu for useful discussions.  This work is supported in part by the Department of Energy under Grant No. DE-FG02-91ER40671.

\bibliographystyle{JHEP}
\bibliography{deform}

\end{document}